# Annular Groove Phase Mask coronagraph in diamond for mid-IR wavelengths: manufacturing assessment and performance analysis

C. Delacroix*[a], P. Forsberg[b], M. Karlsson[b], D. Mawet[c], C. Lenaerts[d],
S. Habraken[a,d], C. Hanot[e], J. Surdej[e], A. Boccaletti[f] and J. Baudrand[f]

[a]HOLOLAB, Université de Liège, 17,B5a Allée du 6 Aout, B-4000 Liège, Belgium
[b]Ångström Laboratory, Uppsala University, Lägerhyddsvägen 1, SE-751 21 Uppsala, Sweden
[c]JPL, California Institute of Technology, 4800 Oak Grove Dr, CA 91109 Pasadena, USA
[d]Centre Spatial de Liège, Avenue du Pré-Aily, B-4031 Liège, Belgium
[e]IAGL, Université de Liège, 17,B5c Allée du 6 Aout, B-4000 Liège, Belgium
[f]LESIA, Observatoire de Paris-Meudon, 5 pl J. Janssen, F-92195 Meudon, France

## ABSTRACT

Phase-mask coronagraphs are known to provide high contrast imaging capabilities while preserving a small inner working angle, which allows searching for exoplanets or circumstellar disks with smaller telescopes or at longer wavelengths. The AGPM (Annular Groove Phase Mask, Mawet et al. 2005[1]) is an optical vectorial vortex coronagraph (or vector vortex) induced by a rotationally symmetric subwavelength grating (i.e. with a period smaller than $\lambda/n$, $\lambda$ being the observed wavelength and $n$ the refractive index of the grating substrate). In this paper, we present our first mid-infrared AGPM prototypes imprinted on a diamond substrate. We firstly give an extrapolation of the expected coronagraph performances in the $N$-band (~10 $\mu m$), and prospects for down-scaling the technology to the most wanted $L$-band (~3.5 $\mu m$). We then present the manufacturing and measurement results, using diamond-optimized microfabrication techniques such as nano-imprint lithography (NIL) and reactive ion etching (RIE). Finally, the subwavelength grating profile metrology combines surface metrology (scanning electron microscopy, atomic force microscopy, white light interferometry) with diffractometry on an optical polarimetric bench and cross correlation with theoretical simulations using rigorous coupled wave analysis (RCWA).

**Keywords:** coronagraphy, annular groove phase mask, vector vortex, achromaticity, ZOGs, subwavelength gratings

## 1. INTRODUCTION

Imaging and characterizing faint sources in the close environment of bright astronomical objects at high angular resolution is one of the most difficult and challenging goals of astronomy. In particular for extrasolar planets, very high rejection ratios are needed, due to the high contrast between the faint companion and its much brighter parent star. Typically, an exoplanet is $10^4 - 10^7$ times fainter than its host star in the thermal infrared ($L$-band, $M$-band, $N$-band, $Q$-band, from 3.5 to 20 $\mu m$). Therefore, such direct exoplanet detection necessitates dedicated instruments such as coronagraphs. The classical Lyot coronagraph (Lyot 1939[2]), considered as an amplitude coronagraph, consists of an opaque spot centered on the optical axis which unfortunately occults quite a significant zone around the star (several $\lambda/D$) including the potential companions behind it. Note that amplitude coronagraphs have also come a long way since Lyot's original opaque spot. Recent evolutions include the apodized Lyot coronagraph (Soummer et al. 2003[3]) and the band-limited Lyot coronagraph (Kuchner et al. 2002[4]). While improving Lyot's coronagraph contrast performance, low throughput and inner working angle remain major limitations of this family of coronagraphs. In order to overcome this inherent disadvantage of amplitude based coronagraph, an alternative approach consists of varying the phase of the incident light instead of blocking it. Such a device is called a phase mask coronagraph (Roddier & Roddier 1997[5], Rouan et al. 2000[6]). It acts in a way that a destructive interference occurs which causes the nulling of the bright source in the relayed pupil downstream of the coronagraph.

*cdelacroix@ulg.ac.be ; www.hololab.ulg.ac.be

Although several extrasolar planets have recently been directly imaged (Fomalhaut, HR 8799, Beta Pictorisb, Kalas 2008[7], Marois *et al.* 2008[8], Lagrange *et al.* 2009[9]), these observations were made under exceptional circumstances: especially large planets (considerably larger than Jupiter), very hot and widely separated from their host star. A couple of months ago, one of us demonstrated that a vector vortex coronagraph (VVC), i.e. a phase mask in which the phase-shift varies azimuthally around the center, could enable small telescopes to directly image exoplanets thanks to its small inner working angle (Mawet *et al.* 2010[10], Serabyn *et al.* 2010[11]). The VVC used in these experiments was made out of liquid crystal polymer (LCP). Here we pursue a different technological route to synthesize the phase ramp. Instead of LCP, we use subwavelength gratings, which are particularly adapted to longer wavelengths. The demand of efficient coronagraphs in the mid-infrared is indeed increasing, following the recent success of high contrast imaging of exoplanets and circumstellar disks in the *L*-band, *N*-band and *Q*-band (e.g. Lagrange *et al.* 2010[12], Moerchen *et al.* 2007[13]).

In this paper, we present the results of our work on the Annular Groove Phase Mask (AGPM) coronagraph, an achromatic VVC proposed in 2005 by our team (Mawet *et al.* 2005[1]). We detail the manufacturing and measurement results obtained with our first mid-infrared AGPM prototype imprinted on a diamond substrate, and its theoretical performances. Finally, we conclude with the perspectives for present and future instruments.

## 2. THEORETICAL BACKGROUND

The Annular Groove Phase Mask (AGPM) coronagraph is a micro-optical element that consists of a concentric circular grating. The grating is subwavelength, preventing it from diffracting light, but forcing it into the zeroth order instead. While transmitting a clean wavefront, the AGPM's artificial birefringence modifies the polarization structure of the beam in a circularly symmetric way that induces a phase ramp in two orthogonal circular polarization components. The embedded phase ramps generate a phase singularity that propagates downstream to create a vortex coronagraph (see Mawet *et al.* 2010[10]).

The geometry of the grooves is defined by their depth $h$, their period $\Lambda$ and their filling factor $F$, which corresponds to the ratio between the width of the lines and the period (see Fig. 1). The profile of the grooves is theoretically supposed to be rectangular. However, the performances of the microfabrication techniques required to fabricate such gratings are limited and the sides of the etched grooves are not perfectly vertical. Thus, another parameter to take into account in order to characterize the geometry of the grating is the slope $\alpha$ of the walls.

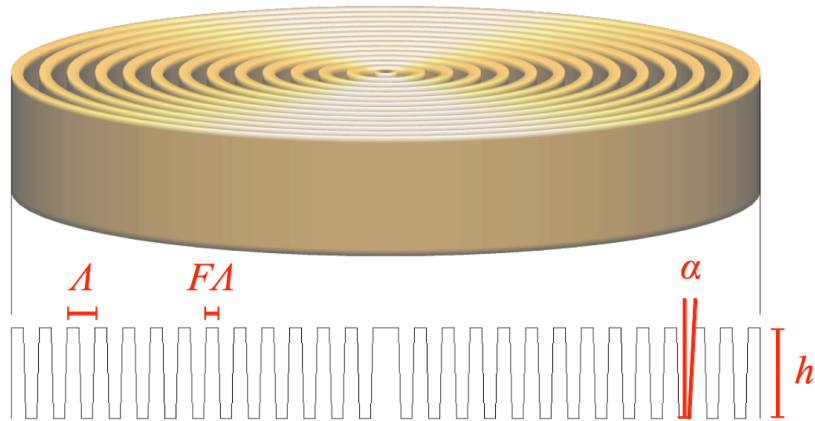

**Figure 1.** Geometry of the Annular Groove Phase Mask: parameters definition.

### 2.1 AGPM coronagraph working principle

The principle of the AGPM is illustrated in Fig. 2. It nominally works with unpolarized natural light. The collimated beam coming from the circular entrance pupil of the telescope converges in the focal plane where the mask is introduced and well-centered, so that the Airy disk i.e. the Fourier transform of the pupil, is focused on the very center of the circular grating. In the focal plane, the mask affects the phase of the beam by a Pancharatnam $4\pi$ phase ramp. In fact, the

AGPM consists of a space-variant subwavelength grating which synthesizes a vectorial optical vortex (vectorial means that the phase shift occurs between vectorial components *s* and *p*). Such a phase ramp corresponds to a topological charge *l* = 2 which analytically leads to a total starlight rejection in the theoretically perfect case (Foo *et al.* 2005[14], Mawet *et al.* 2005[1]). Behind the phase mask, the beam is collimated again inducing another Fourier transform of the electric field. In the next pupil plane, we can notice the effect of the mask which involves a perfect annular symmetry of the rejection around the original pupil. The Lyot stop, slightly undersized compared to the entrance pupil dimension, suppresses the diffracted starlight keeping only the central dark part. Lastly, the light is focused on a detector to produce the final coronagraphic image.

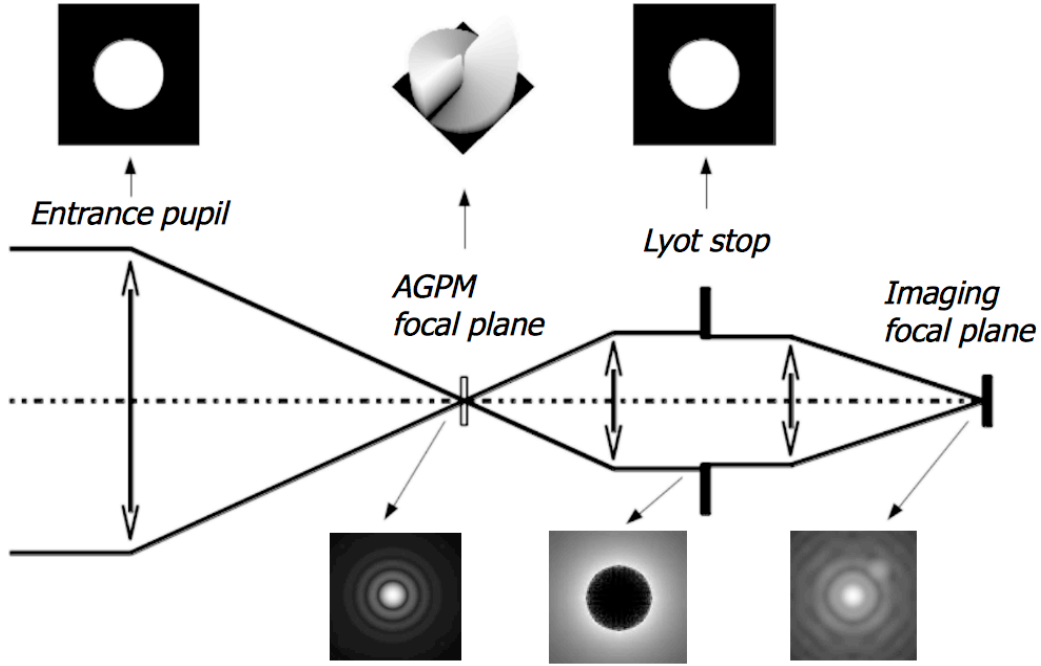

**Figure 2.** AGPM coronagraphic optical bench scheme. Numerical coronagraphic simulation illustrating the diffractive behavior of the AGPM in the *K*-band (Mawet *et al.* 2005[1]).

**2.2 Zeroth Order Gratings (ZOGs)**

Subwavelength gratings are micro-optical structures characterized by a period $\Lambda$ smaller than $\lambda/n$, $\lambda$ being the observed wavelength of the incident light and *n* the refractive index of the grating substrate. Such structures do not diffract light as a classical spectroscopic grating because only zeroth transmitted and reflected orders are allowed to propagate outside the grating, and the incident wavefront is not affected by further aberrations. Therefore, these structures are also called Zeroth Order Gratings (ZOGs). The use of ZOGs in phase mask coronagraphs has been proposed by our team (Mawet *et al.* 2005[15]) as a very promising solution to the recurrent problem of chromaticity. Indeed, other phase mask coronagraphs like the Four Quadrant Phase Mask (FQPM) proposed a few years ago (Rouan *et al.* 2000[6]) and installed on the NAOS-CONICA adaptive optics instrument (Boccaletti *et al.* 2004[16]) at the ESO's Very Large Telescope (VLT), suffer from difficulties to maintain a perfect π phase shift over significantly wide spectral bands such as those usually used as astronomical filters.

The condition for having a ZOG is defined by the grating equation which determines whether a diffraction order propagates or not through the grating:

$$\frac{\Lambda}{\lambda} \leqslant \frac{1}{max(n_i, n_t) + n_i \sin\theta} , \qquad (1)$$

where $\theta$ is the incidence angle and $n_i$ and $n_t$ are respectively the refractive indices of the incident (superstrate) and transmitting (substrate) media. Taking advantage of this property, one can employ these subwavelength gratings to synthesize artificial birefringent achromatic waveplates (Kikuta et al. 1997[17]). Indeed, while propagating in a birefringent medium such as a grating, the incident light sees two different refractive indices, $n_{TE}$ and $n_{TM}$, corresponding to the effective indices of each polarization state TE (transverse electric, or s) and TM (transverse magnetic, or p), and its vectorial components vibrate parallelly or orthogonally to the grating grooves. The differential phase shift between the two polarization components TE (s) and TM (p) is given by the following hyperbolic expression

$$\Delta \Phi_{TE-TM}(\lambda) = \frac{2\pi}{\lambda} h \Delta n_{TE-TM}(\lambda) \qquad (2)$$

where $h$ is the length of the optical (propagation) path through the birefringent medium, which depends on the depth of the grating region and the incidence angle, and $\Delta n_{TE-TM}$ is the form birefringence, i.e. the difference between the refractive indices $n_{TE}$ and $n_{TM}$. The ultimate goal is, by carefully selecting all the grating parameters (geometry, material, incidence), to make the form birefringence proportional to the wavelength over a wide spectral band, in order to ensure the desired constant phase shift, in our case $\pi$ (Mawet et al. 2005[1]).

## 2.3 AGPM versus FQPM

As already mentioned before, the AGPM coronagraph is totally circularly symmetric and actually implements a space-variant ZOG structure, which means that the orientation of the grating lines varies from point to point, unlike the FQPM made out of ZOGs for which the orientation of the grooves is constant in every quadrant. Besides resolving the chromaticity problem of the FQPM, the AGPM also suppresses the phase transitions between adjacent quadrants which induce annoying $\lambda/D$-large "dead zones" that can potentially attenuate a faint companion by up to 4 mag (Riaud et al. 2001[18]), thanks to its circular and continuous spiral phase. Moreover, since it corresponds to a $4\pi$ phase ramp, the AGPM's phase shift distribution in the four cardinal points is analogous to the FQPM. The AGPM can be seen as a FQPM in polarization, and like the FQPM, the theoretical attenuation of the AGPM is infinite in the perfect achromatic and circular filled pupil case.

## 3. NUMERICAL SIMULATIONS AND DESIGN

The scalar theories of diffraction are not efficient in the subwavelength domain. In order to simulate the grating response and to calculate its form birefringence $\Delta n_{TE-TM}$, we must consider the vectorial nature of light. Therefore, we have performed realistic numerical simulations using the Rigorous Coupled Wave Analysis (RCWA, Moharam & Gaylord 1981[19]). The RCWA resolves the Maxwell equations and gives the entire diffractive characteristics of the simulated structure.

## 3.1 Material choice

The use of diamond substrates leads to two major advantages:

- a high refractive index, which is crucial since it is severely linked to the grating aspect ratio (i.e. the ratio between the depth and the feature line) that should be minimized;
- a high transmission over a wide IR spectral band (Bundy 1962[20]), which is likely to suit many instruments like SPHERE on the VLTI, or EPICS and METIS on the future E-ELT (Gilmozzi & Spyromilio 2007[21]).

Previous attempts to fabricate AGPM coronagraphs with other dielectrics (such as silica) have been carried out by our team lately, but where unsuccessful because of the very high required aspect ratios. For the first time, thanks to the expertise of the Ångström Laboratory (Uppsala, Sweden) in the microfabrication of diamond optical components, the targeted specifications have been reached with an acceptable precision, as described further.

## 3.2 Coronagraphic performance simulations in the *N*- and *L*-bands

Here, we present the RCWA results for a subwavelength grating engraved in a diamond substrate. We calculate the null depth *N* as a function of the wavelength, with the optimal geometrical specifications. The null depth quantifies the inherent performance of the coronagraphic device, everything else assumed perfect (no central obscuration or wavefront aberrations, see Mawet *et al.* 2010[22], these proceedings), since it corresponds to the contrast between the darkness of the destructive interference taking place in the relayed pupil plane of the telescope, and the source, i.e. the astronomical bright object. To evaluate the null depth, we take into account the phase errors with respect to $\pi$, that is $\varepsilon(\lambda) = \Delta\Phi_{TE-TM}(\lambda) - \pi$, and the flux ratio $q(\lambda) = \eta_{TE}(\lambda) / \eta_{TM}(\lambda)$ :

$$N(\lambda) = \frac{[1-\sqrt{q(\lambda)}]^2 + \epsilon(\lambda)^2 \sqrt{q(\lambda)}}{[1+\sqrt{q(\lambda)}]^2} \quad . \tag{3}$$

We have chosen to focus on both the *N*-band (9 – 13 *μm*) and the *L*-band (3.5 – 4.1 *μm*), which appear to be very interesting bandwidths for actual and future instruments, especially the *L*-band which has been very successful recently (Lagrange *et al.* 2010[12]). On the other hand, diamond is actually one of the unique solutions for a *N*-band phase mask coronagraph. For this reason, we started the manufacturing using the *N*-band specifications which are defined in Fig. 3. Taking into account the profile metrology (as described in the next section) gives us a very realistic optimal parameters set. By fixing the slope to the measured value ($\alpha = 4.3°$) and the period to the ZOG limit ($\Lambda = 3.78$ *μm* at $\lambda = 9$ *μm* and $\Lambda = 1.47$ *μm* at $\lambda = 3.5$ *μm*), one can easily calculate the optimal value of the two remaining parameters: the filling factor *F* and the depth *h*. In this way, we can extrapolate the performances of the forthcoming mid-IR diamond AGPMs. In Fig. 3, we present the RCWA simulated performances of optimal components (in the *N*- and *L*-bands), realistically producible by our means of microfabrication. The results we obtain are very promising: indeed, the aspect ratio is reduced by 25% compared to our previous attempts with silica. The mean null depths are $\mu \approx 6.1 \times 10^{-3}$ (~$3 \times 10^{-5}$ at $2\lambda/D$, $\lambda$ being the working bandwidth and $D$ the diameter of the telescope) over the whole *N*-band, and $\mu \approx 9.6 \times 10^{-4}$ (~$5 \times 10^{-6}$ at $2\lambda/D$) over the whole *L*-band, which is enough from a scientific standpoint to observe self-luminous planets and disks from the ground, where the main limitation is the turbulence and background noise.

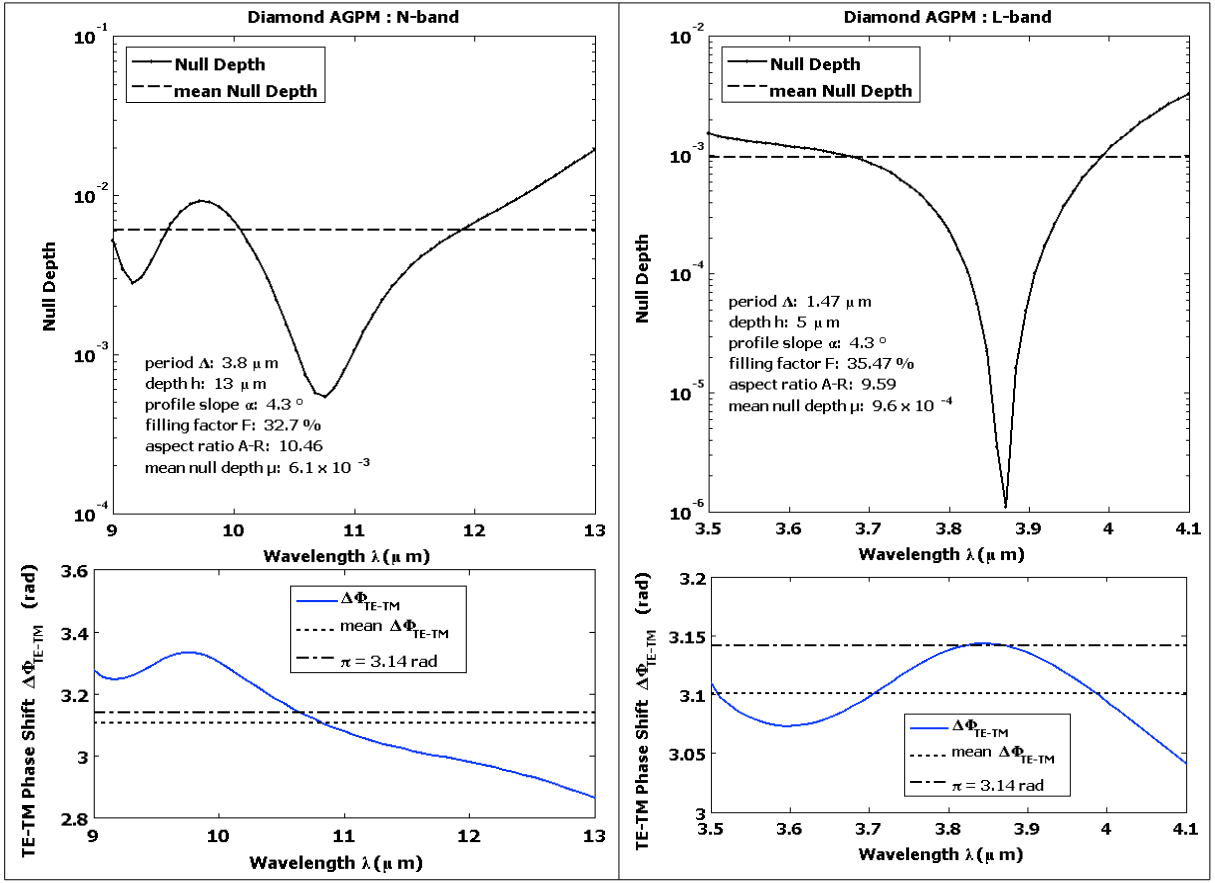

**Figure 3.** Diamond AGPM theoretical performances: null depth and phase shift *vs.* wavelength. Left, for the *N*-band, $\mu \approx 6.1 \times 10^{-3}$ (~3 x 10$^{-5}$ at 2$\lambda$/D). Right, for the *L*-band, $\mu \approx 9.6 \times 10^{-4}$ (~5 x 10$^{-6}$ at 2$\lambda$/D).

### 3.3 Tolerancing

As mentioned before, the shape of the grooves is hard to measure accurately. The average value that we obtain for the slope of the walls is $\alpha = 4.3° \pm 0.3°$ (see the next section). Therefore, we have carried out a tolerancing of the component performances (i.e. the null depth) in function of the slope $\alpha$, for a $\pm 0.3°$ range. We show (see Fig. 4) that we can mitigate the loss of contrast. For this, we compensate for the slope fluctuations by varying the etching depth, which is the easiest parameter to control. We see that the mean null depth, both in the *N*-band and in the *L*-band, does not deviate significantly from its optimal value, while keeping a comfortable aspect ratio (< 11.25).

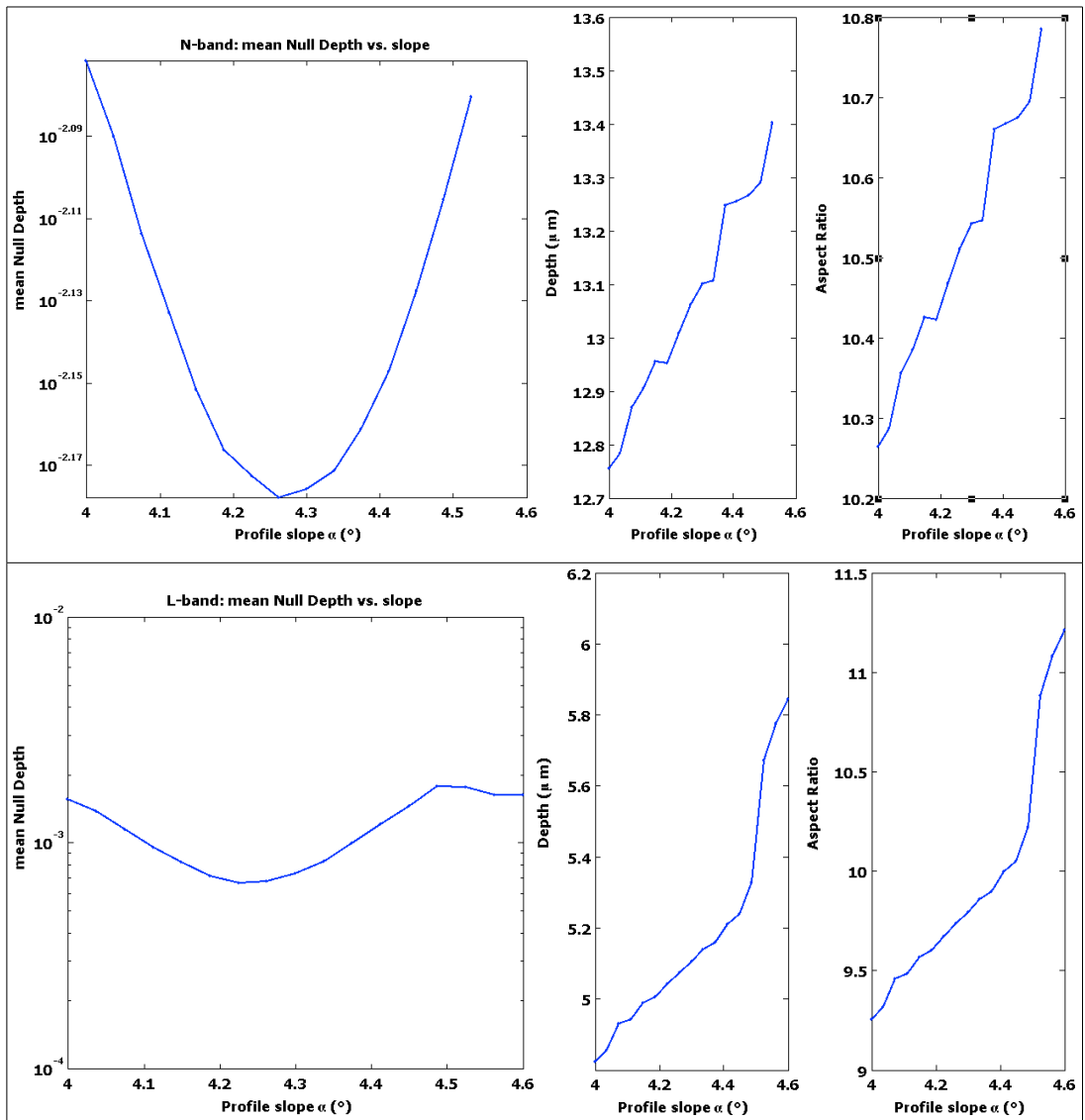

**Figure 4.** Mean null depth vs. slope, while varying the grating depth, for the *N*-band (top) and the *L*-band (bottom).

# 4. MANUFACTURING AND METROLOGY

The prototype manufacturing of the diamond AGPM involves a concerted action with both the Centre Spatial de Liège (Liège, Belgium) and the Ångström Laboratory (Uppsala, Sweden). Currently, we focused our efforts on the mid-IR domain (*N*-band) since it is much easier to achieve. Indeed, we cope with higher periods and lower aspect ratios. Moreover, there is no AR coating needed with this solution (other than on the back side of the substrate), which considerably relieves the microfabrication process. Based on our results of the design optimization study, we elaborated the technology stack and the optimal "recipe" for the microfabrication in order to achieve the required specifications. As expected and thanks to the high refractive index of the diamond that has greatly reduced the etching depth, the diamond etching of the first prototypes has successfully met the specifications: filling factor $F \approx 30\%$, feature line $F\Lambda \approx 1.4~\mu m$, and depth $h \approx 13.5~\mu m$, which corresponds to an aspect ratio $\approx 9.64$.

## 4.1 Manufacturing

The microfabrication has been conducted in the Ångström Laboratory (Uppsala), using circular diamond substrates (diameter = 10 *mm* / thickness = 0.3 *mm*) of optical quality. The fabrication method, based on the laboratory expertise (see Karlsson & Nikolajeff 2003[23], Karlsson *et al.* 2010[24]), was especially developed and optimized for this particular process. It consists of the following steps (see Fig. 5). First, a 500 *nm* aluminum layer is sputtered on top of the diamond substrate, and then a polymer film is spun onto the Al-coating. This polymer, which is a special nano-imprint lithography (NIL) resist, must be spun and baked several times, with more resist added in between, in order to achieve a layer thick enough while minimizing the size of the created edge bead. Indeed, an unwanted edge bead is generated during the spin coating and reduces the effective grating area by about 30-40% in the best case. Next, the surface relief master, previously prepared by direct laser writing (DLW), is imprinted in the NIL-resist following a specific NIL process, and then transferred into the underlying aluminum by use of a reactive ion etching process including inductively coupled plasma (RIE-ICP). The aluminum is etched in a $Cl_2/BCl_3/Ar$ chemistry until it is etched through. The resist is then removed in acetone and the Al-mask carefully inspected by atomic force microscope (AFM) and white light interferometry. Finally, the Al-mask pattern is transferred to the diamond substrate by RIE-ICP in a $O_2/Ar$ chemistry, and the remaining aluminum is removed with solvent.

Let us mention that the master used for the NIL and fabricated by DLW in a 100 *μm* Ni-coated silicon wafer, has been cleaned by sonication in acetone and isopropanol, before being sputtered with chromium (30 *nm*) and gold (100 *nm*). The master was cut into a circle to fit inside the edge bead on the diamond sample. Finally the master was treated with octadecanethiol in isopropanol to reduce sticking problems after imprinting.

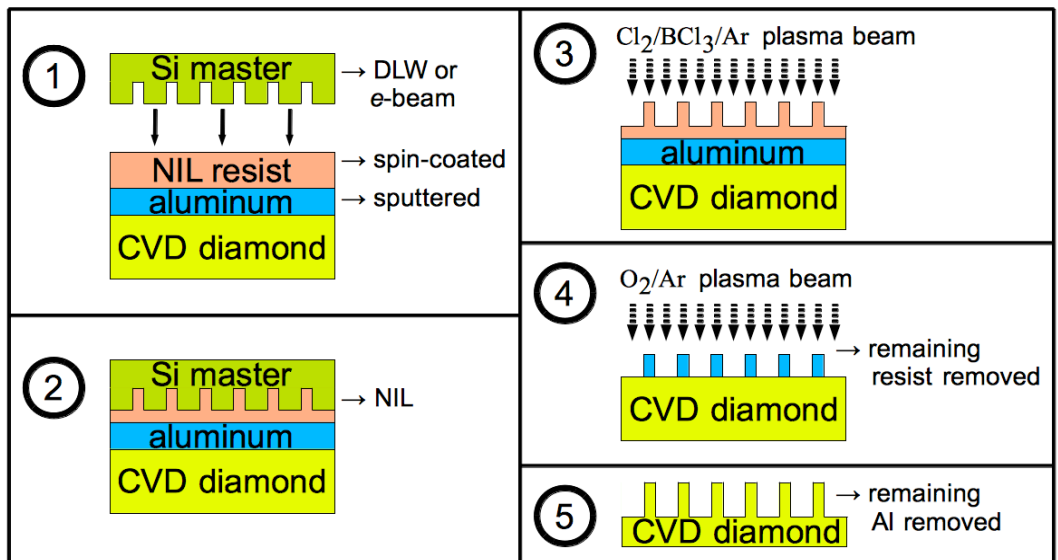

**Figure 5.** Microfabrication process scheme.

## 4.2 Metrology

The subwavelength grating profile metrology has been accomplished in two phases. First, we have carried out a classical surface metrology (see Fig. 6), using scanning electron microscopy (SEM), atomic force microscopy (AFM) and white light interferometry. The measured error on the period $\Lambda$ is less than 5 *nm* which is not affecting significantly the performances at 10 *μm*. However, for future near-IR components with smaller specifications, this precision can be improved by using *e*-beam lithography instead of DLW. Indeed, an *e*-beam mask is theoretically precise within a 1 *nm* range. Concerning the depth *h* of the grooves, it is being well-controlled by the process. The measured error does not exceed several tens of *nm* which is negligible.

The most difficult parameter to control during the fabrication process appears to be the filling factor *F* since it varies in time during the etching. In fact, the deeper the plasma beam penetrates the substrate, the thinner it becomes, so that the sides of the etched grooves are not perfectly vertical. Consequently, the profile of the ridges takes a trapezoidal shape, defined by an additional parameter that is the slope *α* of the walls. Unfortunately, it is not possible to cleave diamond accurately. Thus, we have been investigating a moulding process so as to obtain a replica of the grating that could be cleanly cleaved and precisely measured. This development is currently under assessment in PDMS by the Centre Spatial de Liège. In the meantime, we have conducted a considerable amount of measures in several different places of the component, of both the depth of the grooves (using interferometry) and the slope area size (using optical microscope). Indeed, on the microscope images (Fig. 6), one can distinguish clearly enough the bottom of the ridges from the top. The average value that we obtain for the slope is $\alpha = 4.3° \pm 0.3°$. This value should be shortly confirmed by measures on the forthcoming moulded replica.

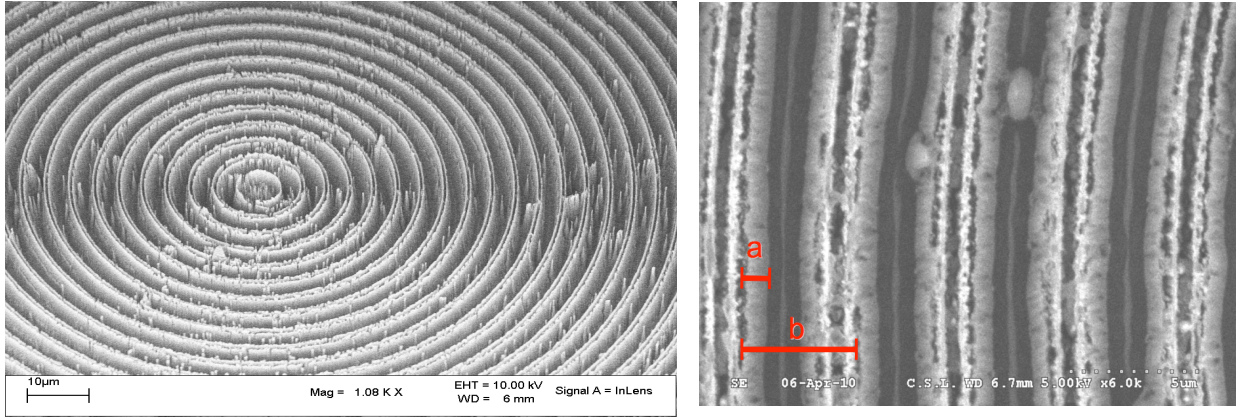

**Figure 6.** Left, SEM image of the first AGPM diamond prototype. Filling factor $F \approx 30\%$, feature line $F\Lambda \approx 1.4$ *μm*, and depth $h \approx 13.5$ *μm*. Right, optical microscope image providing a measure of the slope *α* of the ridge walls: *a* refers to the slope area size, and *b* refers to the period. The average value of the slope is $\alpha = 4.3° \pm 0.3°$.

The second metrology phase consisted of diffractometry. The diffraction efficiencies ($\eta_m$) have been measured on an optical bench and cross correlated with theoretical simulations using rigorous coupled wave analysis (RCWA). To obtain a precise measure with a *N*-band source is naturally very hard. Therefore, we used a visible laser HeNe source (632.82 *nm*) so that we could clearly see the diffraction orders. At this wavelength though, the structure is not subwavelength anymore and one can see many orders ($m = 7$). Besides, the diffusion is very high at 632.82 *nm*. The results we obtain are not accurate enough to deduce the grating profile with precision. For instance, the zeroth order efficiency ($\eta_0$) varies from 10 to 33.3% (see Table 1).

**TABLE 1.** Diffractometry with a HeNe source at 632.82 *nm*.

| $\eta_0$ | $\eta_1$ | $\eta_2$ | $\eta_3$ | $\eta_4$ | $\eta_5$ | $\eta_6$ | $\eta_7$ |
|---|---|---|---|---|---|---|---|
| 10 – 33.3% | 1.3 – 5.4% | 0.9 – 2.5% | 0.8 – 1.8% | 0.7 – 1.2% | 0.6 – 1% | 0.4 – 0.8% | 0.3 – 0.6% |

These ranges are realistic given our assumed profile but the lack of precision does not allow us to extrapolate the shape of the grooves. However, we were able to provide a good measure of the total integrated scattering $TIS = 40 - 80\%$. We can evaluate the roughness with the following expression:

$$R_{RMS}(\lambda) = \frac{\lambda \sqrt{T.I.S.}}{4\pi \cos\theta} \ . \tag{4}$$

For an angle of incidence $\theta = 0°$, we obtain a roughness $R_{RMS} = 32 - 45$ *nm rms* at $\lambda = 632.82$ *nm* [note that 15 *nm* is the *rms* value from the manufacture of the diamond substrate, provided by the supplier *Element6*]. In the mid-IR, such a value is totally acceptable. At 9 *μm* for instance, it corresponds to a $TIS = 0.2 - 0.4\%$ which is negligible.

## 5. CONCLUSION

We have presented the results of the fabrication of the first Annular Groove Phase Mask prototypes, using diamond-optimized microfabrication techniques. This achievement paves the way for the manufacturing of many future AGPM, with a more and more accurate control of the parameters. Knowing the potential performances of this innovative component, this result suggests to expect considerable future progresses in coronagraphy and high dynamic range imaging.

Despite the drastic constraints related to the etching processes, and the particularly high and unusual aspect ratio with regard to diamond microfabrication capabilities, we managed to achieve a result close to the desired specifications. Through a range of metrology techniques, we have characterized the sublambda structure profile. Although the components are not usable for coronagraphic applications yet, the microfabrication process is well known and reproducible.

We extrapolate the expected performances of the forthcoming mid-IR diamond AGPMs, not only in the *N*-band (~10 *μm*) but also in the most wanted *L*-band (~3.5 *μm*). Indeed, a French team has recently been able to follow the motion of an exoplanet by direct imaging (Lagrange *et al.* 2010[12]) in this bandwidth. The instrument used for this exploit (NAOS-CONICA at the VLT) could be a very appropriate candidate to welcome a *L*-band diamond AGPM. The manufacturing of new components is underway. Predicted performances should be sufficient to characterize the component on a coronagraphical bench, at the LESIA (in Paris). This will be the subject of a forthcoming paper.

## ACKNOWLEDGEMENT

We gratefully acknowledge the help of Prof. Markku Kuittinen, Ismo Vartianen and Kari Leinonen (University of Joensuu, Finland) with lithography. C. Delacroix acknowledges the financial support of the Belgian "Fonds pour la formation à la Recherche dans l'Industrie et dans l'Agriculture". The authors from the Liège University acknowledge support from the Communauté française de Belgique - Actions de recherche concertée - Académie universitaire Wallonie-Europe.

## REFERENCES

[1] Mawet, D., Riaud, P., Absil, O. and Surdej, J., "Annular Groove Phase Mask Coronagraph", *ApJ* **633**, 1191-1200 (2005).
[2] Lyot, B., "The study of the solar corona and prominences without eclipses (George Darwin Lecture, 1939)", *MNRAS* **99**, 580-594 (1939).
[3] Soummer, R., Aime, C. and Falloon, P.-E., "Stellar coronagraphy with prolate apodized circular apertures", *A&A* **397**, 1161-1172 (2003).
[4] Kuchner, M. and Traub, W., "A Coronagraph with a Band-limited Mask for Finding Terrestrial Planets", *ApJ* **570**, 900-908 (2002).
[5] Roddier, F. and Roddier, C., "Stellar Coronograph with Phase Mask", *PASP* **109**, 815-820 (1997).

[6] Rouan, D., Riaud, P., Boccaletti, A., Clénet, Y. and Labeyrie, A., "The Four-Quadrant Phase-Mask Coronagraph. I. Principle", *PASP* **112**, 1479-1486 (2000).

[7] Kalas, P., Graham, J., Chiang, E., et al., "Optical Images of an Exosolar Planet 25 Light-Years from Earth", *Science* **322**, 1345-1348 (2008).

[8] Marois, C., Macintosh, B., Barman, T., et al., "Direct Imaging of Multiple Planets Orbiting the Star HR 8799", *Science* **322**, 1348-1352 (2008).

[9] Lagrange, A.-M., Gratadour, D., Chauvin, G., et al., "A probable giant planet imaged in the *β* Pictoris disk. VLT/NaCo deep L'-band imaging", *A&A* **493**, L21-L25 (2009).

[10] Mawet, D., Serabyn, E., Liewer, K., Burruss, R., Hickey, J. and Shemo, D., "The Vector Vortex Coronagraph: Laboratory Results and First Light at Palomar Observatory", *ApJ* **709**, 53-57 (2010).

[11] Serabyn, E., Mawet, D. and Burruss, R., "An image of an exoplanet separated by two diffraction beamwidths from a star", *Nature* **464**, 1018-1020 (2010).

[12] Lagrange, A.-M., Bonnefoy, M., Chauvin, G., et al., "A Giant Planet Imaged in the Disk of the Young Star *β* Pictoris", *Science* (2010); doi: 10.1126/science.1187187.

[13] Moerchen, M., Telesco, C., De Buizer, J., Packham, C. and Radomski, J., "12 and 18 μm Images of Dust Surrounding HD 32297", *ApJ* **666**, L109-L112 (2007).

[14] Foo, G., Palacios, D. and Swartzlander, G., Jr., "Optical vortex coronagraph", *Opt. Lett.* **30**, 3308-3310 (2005).

[15] Mawet, D., Riaud, P., Surdej, J. and Baudrand, J., "Subwavelength surface-relief gratings for stellar coronagraphy", *Appl. Opt.* **44**, 7313-7321 (2005).

[16] Boccaletti, A., Riaud, P., Baudoz, P., et al., "The Four-Quadrant Phase Mask Coronagraph. IV. First Light at the Very Large Telescope", *PASP* **116**, 1061-1071 (2004).

[17] Kikuta, H., Ohira, Y. and Iwata, K., "Achromatic quarter-wave plates using the dispersion of form birefringence", *Appl. Opt.* **36**, 1566-1572 (1997).

[18] Riaud, P., Boccaletti, A., Rouan, D., Lemarquis, F. and Labeyrie, A., "The Four-Quadrant Phase-Mask Coronagraph. II. Simulations", *PASP* **113**, 1145-1154 (2001).

[19] Moharam, M. and Gaylord, T., "Diffraction analysis of dielectric surface-relief gratings," *JOSA* **72**, 1385-1392 (1982).

[20] Bundy, F., "Melting Point of Graphite at High Pressure: Heat of Fusion", *Science* **137**, 1055-1057 (1962).

[21] Gilmozzi, R. and Spyromilio, J., "The European Extremely Large Telescope (E-ELT)", *The Messenger* **127**, 11-19 (2007).

[22] Mawet, D. and Serabyn, E., "The vector vortex coronagraph: analysis of sensitivity to low-order aberrations, central obscuration and chromatism", *Proc. SPIE* **7739** (2010).

[23] Karlsson, M. and Nikolajeff, F., "Diamond micro-optics: microlenses and antireflection structured surfaces for the infrared spectral region", *Opt. Express* **11**, 502-507 (2003).

[24] Karlsson, M., Vartianen, I., Kuittinen, M. and Nikolajeff, F., "Fabrication of sub-micron high aspect ratio diamond structures with nanoimprint lithography", *Microelectron. Eng.* (2010); doi:10.1016/j.mee.2009.12.085.